\documentclass[12pt,a4paper]{article}
 
\usepackage{jheppub}
\usepackage[usenames,dvipsnames]{xcolor}
\usepackage{amssymb} 
\usepackage{amsmath}
\usepackage{mathtools}
\usepackage{amsfonts}    
\usepackage{dsfont}
\usepackage{pdfpages}
\usepackage{verbatim}
\usepackage{stmaryrd}
\usepackage{graphicx}
\usepackage{braket}
\usepackage{cancel}
\usepackage{tensor}
\usepackage{mathrsfs}
\usepackage{textgreek} 
\usepackage[mathscr]{euscript}
\usepackage[smalltableaux]{ytableau}
\ytableausetup{centertableaux}
\usepackage{tikz}
\usetikzlibrary{decorations.pathreplacing, decorations.markings,calc,shapes.misc,decorations.pathmorphing,patterns.meta, math}
\usepackage{slashed}
\usepackage{datetime}
\usepackage{hyperref}
\hypersetup{
    pdfencoding=unicode,
	colorlinks=true,
	urlcolor=Maroon,
	linkcolor=RoyalBlue,
	citecolor=Maroon,
	pdftitle={$2$-Split of Form Factors via BCFW Recursion Relation},
	pdfauthor={Liang Zhang},
	pdfdisplaydoctitle=true,
	pdfstartview=FitH,
	linktocpage=true
}

\usetikzlibrary{shapes}

\newcommand{\ba}{\begin{align}}
\newcommand{\ea}{\end{align}}
\newcommand{\be}{\begin{equation}}
\newcommand{\ee}{\end{equation}}
\newcommand{\bea}{\begin{eqnarray}}
\newcommand{\eea}{\end{eqnarray}}

\def\bd{\begin{tikzpicture}}
\def\ed{\end{tikzpicture}}

\allowdisplaybreaks

\def\XXint#1#2#3{{\setbox0=\hbox{$#1{#2#3}{\int}$}
     \vcenter{\hbox{$#2#3$}}\kern-.5\wd0}}

\definecolor{light-gray}{gray}{0.75}
\newcommand{\nn}{\nonumber \\}

\newcommand\Tr{\mathrm{Tr}}

\renewcommand{\ge}{\geqslant}

\title{$2$-Split of Form Factors via BCFW Recursion Relation}
\author{Liang Zhang}
\emailAdd{liangzhang@csrc.ac.cn}
\affiliation{Beijing Computational Science Research Center, Beijing 100084, China}

\abstract{Recently, \cite{Cao:2025hio} demonstrated the $2$-split for form factor under specific kinematic constraints. This factorization is analogous to that observed in scattering amplitudes. A key consequence of this structure is the presence of hidden zeros, where the form factors vanish on specific kinematic loci. We first establish these zeros and a new zero for the form factors of the composite operators ${\cal O} =\frac{1}{2}\Tr((\partial \phi)^2) + \Tr(\phi^3)$ and ${\cal O} = \Tr(F^2)$, and then employ an inductive proof based on the BCFW recursion relation to prove the $2$-split factorization for any number of external particles.
} 

\begin{document}

\setcounter{tocdepth}{3}
\maketitle
\setcounter{page}{2}

\section{\label{sec:intro}Introduction}

The study of scattering amplitudes has undergone a paradigm shift over the past decades. Recently, a remarkable new structure known as hidden zeros \cite{Arkani-Hamed:2023swr}—the vanishing of various tree-level amplitudes at specific loci in kinematic space—has been discovered. This phenomenon has generated considerable interest and inspired extensive research from multiple perspectives \cite{Rodina:2024yfc,Li:2024qfp,Zhang:2024iun,Zhou:2024ddy,Zhang:2024efe,Bartsch:2024amu,Li:2025suo,Huang:2025blb,Backus:2025hpn,De:2025bmf,Jones:2025rbv}. Closely related to this is a novel factorization property, the $2$-split \cite{Cao:2024gln,Cao:2024qpp}, which has likewise attracted substantial attention \cite{Arkani-Hamed:2024fyd,Zhang:2024iun,Zhou:2024ddy,Zhang:2024efe,Feng:2025ofq,Feng:2025dci}. The $2$-split arises when a particular constraint is imposed, under which the amplitude factorizes into two amputated currents.

An important question is whether these remarkable properties extend beyond scattering amplitudes to other fundamental observables. Form factors, which describe the coupling of an off-shell local operator to on-shell asymptotic states, provide a crucial bridge between on-shell and off-shell physics. They are defined as
\begin{align}
F_n^{{\cal O}}(1,2,\ldots,n;q)\equiv \int d^Dx,e^{iq\cdot x}\braket{0|{\cal O}(x)|1,2,\ldots,n}\,,
\end{align}
where $q=\sum_{i=1}^n p_i$. Compared with general correlation functions, form factors are structurally simpler yet retain rich information about the operator. Understanding their analytic properties is therefore essential both for theoretical consistency and for phenomenological applications.

Indeed, the extension of the $2$-split to form factors has been demonstrated in \cite{Cao:2025hio}, where it was shown that for the operators ${\cal O}=\tfrac{1}{2}\mathrm{Tr}[(\partial \phi)^2]+\mathrm{Tr}(\phi^3)$ and ${\cal O}=\mathrm{Tr}(F^2)$, string-theoretic integral representations exhibit the same factorization structure. Specifically, in certain kinematic regions, these form factors decompose into the product of a lower-point off-shell amplitude and a lower-point form factor.

The on-shell program has achieved substantial progress in the study of form factors \cite{Brandhuber:2010ad,Bork:2010wf,Brandhuber:2011tv,Brandhuber:2012vm,Brandhuber:2014ica,Boels:2012ew,Bork:2014eqa,Frassek:2015rka,Bork:2016hst,Bork:2016xfn,Yang:2016ear,Bork:2017qyh,Bianchi:2018peu,Nandan:2018hqz,Yang:2019vag,Lin:2022jrp,Lin:2023rwe}. A cornerstone of this framework is the Britto–Cachazo–Feng–Witten (BCFW) recursion relation \cite{Britto:2004ap,Britto:2005fq}, which enables tree-level amplitudes to be constructed recursively from simpler, lower-point inputs. In this work, we extend the use of BCFW recursion relation to establish the $2$-split of tree-level form factors for specific composite operators in scalar and gauge theories. The proof closely parallels the amplitude case studied in \cite{Feng:2025ofq}, but with a crucial difference: while the zeros of amplitudes naturally induce zeros for form factors, an additional hidden zero unique to form factors is required in order to establish the $2$-split.

The structure of this paper is as follows. In Section \ref{sec:review}, we present a brief review of the $2$-split behavior. In Section \ref{sec:zeros}, we identify the hidden zeros and a new zero relevant to the form factors of ${\cal O}=\tfrac{1}{2}{\rm Tr}[(\partial \phi)^2]+{\rm Tr}(\phi^3)$ and ${\cal O}={\rm Tr}(F^2)$. In Section \ref{sec:split}, we employ these results within an inductive proof based on BCFW recursion relation, thereby establishing the $2$-split factorization for arbitrary multiplicity. We conclude in Section \ref{sec:con} with a brief summary and outlook.

\section{Review}
\label{sec:review}
Let us first briefly review the so-called $2$-split behavior discovered in \cite{Cao:2024gln,Cao:2024qpp}, where certain amplitudes factorize into two currents under specific kinematic constraints. Consider the ${\rm Tr}(\phi^3)$ theory as an example. Select three distinguished particles $i,j,k$ and partition the remaining external legs into two disjoint sets $A$ and $B$ such that $A\cup B=\{1,\ldots,n\}\setminus\{i,j,k\}$. Impose the constraint
\begin{align}\label{eq:amp2splitcons}
	s_{a,b}=0\,, \quad \forall a\in A, b\in B\,,
\end{align}
where the Mandelstam invariants are $s_{a,\ldots,m}=(p_a+\ldots +p_m)^2$. Under this condition, the amplitude ${\cal A}^{\phi^3}_n(i,A,j,B(k))$ exhibits a $2$-split factorization:
\begin{align}\label{eq:ampphi3split}
	{\cal A}^{\phi^3}_n(i,A,j,B(k))\to {\cal J}^{\phi^3}_{|A|+3}(i,A,j;\kappa){\cal J}^{\phi^3}_{|B|+3}(j,B,i;\kappa')\,,
\end{align}
where $B(k)\equiv B\cup \{k\}$ preserves the ordering\footnote{For instance, $B=\{1,2,4\},k=3$, the $B(k)=\{1,2,3,4\}$.}.  The semicolon indicates the off-shell leg, not shoen in the ordering, and $\kappa,\kappa'$ take the position of the leg $k$\footnote{For example, if $k=n$, then $(i,A,j;\kappa)=(i,A,j,\kappa)$ and $(j,B,j;\kappa')=(j,\ldots,n-1,\kappa',1,\ldots,i)$.}.  

As an illustration, the $5$-point amplitude is
\begin{align}
	{\cal A}^{\phi^3}(2,3,4,5,1)=\frac{1}{s_{1,2}}\frac{1}{s_{1,2,3}}+\frac{1}{s_{2,3}}\frac{1}{s_{1,2,3}}+\frac{1}{s_{2,3}}\frac{1}{s_{2,3,4}}+\frac{1}{s_{1,2}}\frac{1}{s_{3,4}}+\frac{1}{s_{3,4}}\frac{1}{s_{2,3,4}}\,.
\end{align}
Setting $i=2, j=4, k=5$, and applying $s_{a,b}=s_{3,1}=0$, the amplitude factorizes as
\begin{align}
	{\cal A}^{\phi^3}(2,3,4,5,1) \to& \left(\frac{1}{s_{2,3}}+\frac{1}{s_{3,4}}\right)\, \left(\frac{1}{s_{1,2}}+\frac{1}{s_{1,5}}\right)\nn
	\equiv& {\cal A}^{\phi^3}(2,3,4;\kappa) {\cal A}^{\phi^3}(4,1,2;\kappa')\,.
\end{align}

For Yang-Mills (YM) amplitudes, additional polarization constraints are required:
\begin{align}\label{eq:ym2splitcons}
	\epsilon_a \cdot \epsilon_{b'}=p_a \cdot \epsilon_{b'}=\epsilon_a \cdot p_b=0 \,,
\end{align}
where $b'\in B'=B\cup\{i,j,k\}$. Under the constraints \eqref{eq:amp2splitcons} and \eqref{eq:ym2splitcons}, the YM amplitude factorizes as
\begin{align}
	{\cal A}^{\rm YM}(i,A,j,B(k))\to {\cal J}^{\rm mixed}_{|A|+3}(i^{\phi},A,j^{\phi};\kappa^{\phi})\,{\cal J}^{\rm YM}_{|B|+3}(j,B,i;\kappa')_{\mu}\epsilon_k^{\mu}\,,
\end{align}
where the pure gluon current contracts with polarization $\epsilon_k$.  As a simple example, consider the $4$-point Yang–Mills amplitude with $i=1$, $j=3$, $k=4$, and $A={2}$. Under the $2$-split kinematic and polarization constraints, the amplitude factorizes as
\begin{align}
	{\cal A}^{\rm YM}(1,2,3,4)\to  &\left(\frac{p_1\cdot \epsilon _2}{s_{12}}-\frac{p_3\cdot \epsilon _2}{s_{2,3}}\right)\left(\epsilon _1\cdot \epsilon _3 p_{3,\mu} + p_1\cdot \epsilon _3 \epsilon _{1,\mu}- p_3\cdot \epsilon _1 \epsilon _{3,\mu}\right)\epsilon _4^{\mu}\nn
	&\equiv {\cal J}^{\rm mixed}(1^{\phi},2,3^{\phi};\kappa^{\phi}){\cal J}_{\mu}^{\rm YM}(1,3;\kappa')\epsilon_4^{\mu}\,.
\end{align}
More examples and effective theory amplitudes can be found in \cite{Cao:2024gln,Cao:2024qpp}.

\section{Hidden zeros}
\label{sec:zeros}
We employ the BCFW recursion relation to prove the $2$-split behavior of form factors. A key question is whether the shifted form factor vanishes as $z \to \infty$ under the BCFW shift,
\begin{align}\label{eq:bcfwshift}
	p_i\to\hat{p}_i= p_i+z r, ~~~~~~ p_j\to\hat{p}_j= p_j+z r
\end{align}
with $r^2=r\cdot p_i=r\cdot p_j=0$. 

We decompose the operator ${\cal O}$ into two parts: a kinematic term ${\cal O}_{\rm kin}$ and an interaction term ${\cal O}_{\rm int}$. The insertion of ${\cal O}_{\rm int}$ makes the form factor behave analogously to the corresponding amplitude, while the essential difference between a form factor and a pure amplitude arises from ${\cal O}_{\rm kin}$. This term introduces a non-trivial numerator but also provides an additional propagator.  Consequently, the relation between a form factor and the corresponding amplitude takes the form
\begin{align}
	F_n^{\cal O}=F^{{\cal O}_{\rm kin}}_n+F^{{\cal O}_{\rm int}}_n=\sum_i \frac{V^{{\cal O}_{\rm kin}}}{(p_i+q)^2}{\cal A}_n+ C {\cal A}_n\,,
\end{align}
where $C$ is a constant, and the summation runs over all external and internal legs at which the operator ${\cal O}_{\rm kin}$ is inserted. Therefore, our scaling analysis reduces to examining the large-$z$ behavior of the product formed by the additional numerator and the propagator arising from the insertion of ${\cal O}_{\rm kin}$.

As an example, consider the operator ${\cal O} = {\cal L}_{\phi} \equiv \frac{1}{2}{\rm Tr}\left[(\partial \phi)^2\right] + {\rm Tr}(\phi^3)$. Here ${\cal O}_{\rm kin} = \frac{1}{2}{\rm Tr}\left[(\partial \phi)^2\right]$ and ${\cal O}_{\rm int} = {\rm Tr}(\phi^3)$. The $3$-point form factor is given by
\begin{align}
	F^{{\cal L}_{\phi}}_3(1,2,3;q)=&F^{{\cal O}_{\rm kin}}_3(1,2,3;q)+F^{{\cal O}_{\rm int}}_3(1,2,3;q)\nn
	=&\frac{s_{1,2}+s_{1,3}}{s_{2,3}}+\frac{s_{1,2}+s_{2,3}}{s_{1,3}}+\frac{s_{1,3}+s_{2,3}}{s_{1,2}}+2\nn
	=&\frac{(s_{1,2}+s_{2,3})(s_{1,2}+s_{1,3})(s_{1,3}+s_{2,3})}{(s_{1,2}s_{1,3}s_{2,3})}\,.
\end{align}

For the amplitude ${\cal A}^{\phi^3}$, since the $i$ and $j$ are non-adjacent, the large $z$ scaling is at least $z^{-1}$.  For the YM amplitude ${\cal A}^{\rm YM}$, the scaling is $z^{-2}$ \cite{Arkani-Hamed:2008bsc}. We now examine whether the product of the additional vertex $V^{\cal O}(m,n;q)$ and the extra propagator alters this scaling and produces a non-vanishing boundary term. 

For the operator ${\cal L}_{\phi}$, the extra vertex behave as $V^{{\cal L}_{\phi}}(m,n;q)\propto p_m \cdot q\sim z$, while the additional propagator  scales as $z^{-1}$ for large $z$. Hence, the overall scaling of $F_n^{{\cal L}_\phi}$ is $z^{-1}$, identical to ${\cal A}^{\phi^3}$. For ${\cal O}=\rm Tr(F^2)$, the additional vertex is $V_{\mu\nu}^{{\rm Tr}(F^2)}(m,n;q)=p_m\cdot q g_{\mu\nu}-p_{m,\mu}q_{\nu}$, which scales as $z^2$ at most when contracted with a momentum $p$ or polarization $\epsilon$. The extra propagator scales as $z^{-1}$, while ${\cal A}^{\rm YM}$ scales as $z^{-2}$. Therefore, the overall scaling of $F_n^{{\rm Tr}(F^2)}$ is again $z^{-1}$. Consequently, for the form factors considered in this paper, the boundary term vanishes.

Analogous to the BCFW recursion relation in \cite{Feng:2025ofq}, if the form factor possesses hidden zeros, the recursion formula can be greatly simplified. Consider the operator ${\cal L}_{\phi}$ with external particles taken to be scalars in the $\mathrm{Tr}(\phi^3)$ theory. Analogous to the amplitude, the form factor $F^{{\cal L}_{\phi}}_n(1,\ldots,n;q)$ exhibits a $2$-split factorization:
\begin{align}\label{eq:phi3split}
	F^{{\cal L}_{\phi}}_n(i,A,j,B(k);q)\to {\cal A}^{\phi^3}_{|A|+3}(i,A,j;\kappa)F^{{\cal L}_{\phi}}_{|B|+3}(j,B,i;\kappa',q)\,.
\end{align}
under these kinematic conditions
\begin{align}\label{eq:2splitcons}
    s_{a,b}=s_{a,q}=0\,, \quad \forall a\in A, b\in B\,,
\end{align}
where $s_{a,q}=2 p_a\cdot q$. 

The resulting amplitude is known to exhibit hidden zeros \cite{Arkani-Hamed:2023swr}, namely
\begin{align}
    {\cal A}^{\phi^3}_{|A|+3}(i,A,j;\kappa)\xrightarrow{s_{a,\kappa}=0} 0\,,
\end{align}
where momentum conservation implies $p_{\kappa}=p_B+p_k-q$, so that $s_{a,\kappa}=s_{a,k}$. Consequently, the form factor also vanishes under the same conditions:
\begin{align}\label{eq:ffzero}
    F^{{\cal L}_{\phi}}_n(i,A,j,B(k);q)\xrightarrow{\eqref{eq:2splitcons},s_{a,k}=0\,} 0\,.
\end{align}

In addition, the form factor possesses other hidden zeros not directly inherited from the amplitude. Specifically, we claim that
\begin{align}\label{eq:ffzero2}
    F^{{\cal L}_{\phi}}_n(i,\alpha,j;q)\xrightarrow{s_{a,q}=0,\forall a\in \alpha} 0\,.
\end{align}
This property can be proven inductively. The base case is the three-point form factor,
\begin{align}
    F^{{\cal L}_{\phi}}_3(1,2,3;q)=\frac{(s_{1,2}+s_{2,3})(s_{1,2}+s_{1,3})(s_{1,3}+s_{2,3})}{(s_{1,2}s_{1,3}s_{2,3})}\,.
\end{align}
This expression clearly vanishes when the condition $s_{2,q}=s_{1,2}+s_{2,3}=0$ is applied.

Assuming \eqref{eq:ffzero2} holds for form factors with fewer than $n$ points, we can establish it for $F^{{\cal L}_{\phi}}_n$ using the BCFW recursion relation.  Under the BCFW shift \eqref{eq:bcfwshift}, We apply the BCFW recursion relation in the form factor \cite{Brandhuber:2010ad}:
\begin{align}
    F^{{\cal L}_{\phi}}_n(i,\alpha,j;q)=\sum_{I}F^{{\cal L}_{\phi}}_{L}(\hat{i},\alpha_L,\hat{I};q)\frac{1}{P_I^2}{\cal A}^{\phi^3}_{R}(-\hat{I},\alpha_R,\hat{j})+{\cal A}^{\phi^3}_{L}(\hat{i},\alpha_L,\hat{I})\frac{1}{P_I^2}F^{{\cal L}_{\phi}}_{R}(-\hat{I},\alpha_R,\hat{j};q)\,.
\end{align}
In the kinematic limit where $s_{a,q}=0$ for all $a\in\alpha$, every term in the BCFW sum involves a lower-point form factor $F^{{\cal L}_{\phi}}_{L}$ or $F^{{\cal L}_{\phi}}_{R}$ evaluated on a subset of these legs. By the induction hypothesis, these lower-point form factors vanish. Consequently, the entire right-hand side is zero. This completes the proof, confirming that \eqref{eq:ffzero2} is valid for all $n\ge 3$.

Similarly, YM form factor for the operator ${\cal O}=\rm Tr(F^2)$ possesses hidden zeros. The YM form factor also exhibits a $2$-split factorization on a different kinematic locus:
\begin{align}\label{eq:ymsplit}
 F^{\rm Tr(F^2)}_{n}(i,A,j,B(k);q)\,\xrightarrow[]{}\,{\cal J}^{\rm mixed}_{|A|+3}(i^{\phi},A,j^{\phi};\kappa^{\phi})\,F^{{\rm Tr}(F^2)}_{|B|+3}(j,B,i;\kappa',q)_{\mu}\epsilon_k^{\mu}\,,
\end{align} 
under the kinematic constraint of \eqref{eq:2splitcons} and the polarization constraint:
\begin{align}\label{eq:polarcon}
    \epsilon_a\cdot \epsilon_{b'}=\epsilon_a\cdot p_b=\epsilon_{b'}\cdot p_a=\epsilon_a\cdot q=0\,, \quad \forall a\in A, b\in B, b'\in B\cup\{i,j,k\}\,.
\end{align}
Here, ${\cal J}^{\rm mixed}_{|A|+3}$ is a current with $|A|$ gluons and $3$ $\phi$'s. The resulting current is known to have hidden zeros \cite{Arkani-Hamed:2023swr}, vanishing when $s_{a,\kappa}=\epsilon_a\cdot p_{\kappa}=0$. Given the momentum definitions, this condition becomes $s_{a,k}=\epsilon_a\cdot p_k=0$. This implies that the entire form factor vanishes on this locus:
\begin{align}
    {\cal J}^{\rm mixed}_{|A|+3}(i^{\phi},A,j^{\phi};\kappa^{\phi})\xrightarrow{s_{a,\kappa}=\epsilon_a\cdot p_{\kappa}=0} 0\,,
\end{align}
Thus, the YM form factor vanishes with the same constraint:
\begin{align}\label{eq:ymhiddenzero}
    F^{\rm Tr(F^2)}_{n}(i,A,j,B(k);q)\xrightarrow[\eqref{eq:2splitcons}\eqref{eq:polarcon}]{s_{a,k}=\epsilon_a\cdot p_k=0}0\,.
\end{align}

We claim
\begin{align}\label{eq:ymffzero2}
    F^{{\cal L}_{\phi}}_n(i,\alpha,j;q)\xrightarrow{s_{a,q}=\epsilon_a\cdot q=0,\forall a\in \alpha} 0\,.
\end{align}
We can prove it by induction. The simplest case is $F^{\Tr(F^2)}_3(1,2,3;q)$, given in \cite{Dong:2022bta}:
\begin{align}
    F^{\Tr(F^2)}_3(1,2,3;q)=\frac{1}{s_{1,2}}\left({\rm tr}(2,3)p_2\cdot\epsilon_1-{\rm tr}(1,3)p_1\cdot \epsilon_2+2{\rm tr}(1,2,3)\right)+{\rm cylic}(1,2,3)\,,
\end{align}
where ${\rm tr}(\alpha_1,\ldots,\alpha_m)\equiv (f_{\alpha_1})_{\mu_1}^{\mu_2}\cdots(f_{\alpha_m})_{\mu_m}^{\mu_1}$ and $f_i^{\mu\nu}=p_i^{\mu}\epsilon_i^{\nu}-p_i^{\mu}\epsilon_i^{\nu}$. This expression can be verified to vanish under the conditions $s_{2,q}=s_{1,2}+s_{2,3}=0$ and $\epsilon_2\cdot q=\epsilon_2\cdot p_1+\epsilon_2\cdot p_3=0$. 

The BCFW recursion relation for the YM form factor is  \cite{Brandhuber:2010ad}
\begin{align}
    F^{\Tr(F^2)}_n(i,\alpha,j;q)=\sum_{I}\sum_{h}&\Big(F^{\Tr(F^2)}_{L}(\hat{i},\alpha_L,\hat{I};q)\frac{1}{P_I^2}{\cal A}^{\phi^3}_{R}(-\hat{I},\alpha_R,\hat{j})\nn
    &+{\cal A}^{\phi^3}_{L}(\hat{i},\alpha_L,\hat{I})\frac{1}{P_I^2}F^{\Tr(F^2)}_{R}(-\hat{I},\alpha_R,j;q)\Big)\,,
\end{align}
where the $\sum_h$ is sum over the helicities of the internal gluon $I$. As before, by the induction hypothesis, the lower-point form factors on the right-hand side, $F^{\Tr(F^2)}_{L}$ and $F^{\Tr(F^2)}_{R}$, vanish in the specified kinematic limit. Thus,  \eqref{eq:ymffzero2} holds for all $n\ge 3$.

\section{$2$-split}
\label{sec:split}
\subsection{${\cal O}=\frac{1}{2}\rm Tr\left((\partial \phi)^2\right)+\rm Tr(\phi^3)$}

\begin{figure}[htb]
  \centering
  \includegraphics[width=0.4\linewidth]{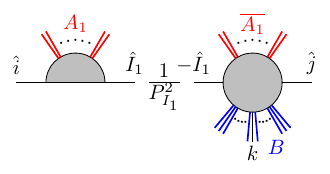}
   \includegraphics[width=0.4\linewidth]{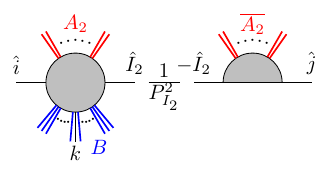}
   \includegraphics[width=0.4\linewidth]{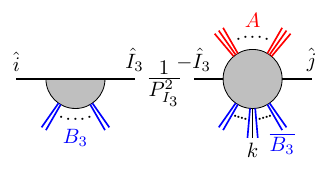}
   \includegraphics[width=0.4\linewidth]{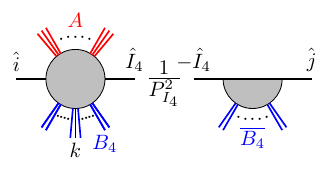}
  \caption{The four non-vanishing BCFW diagrams that contribute to the form factors.}
  \label{fig:bcfw}
\end{figure}

We will now prove the 2-split behavior of the form factor, \eqref{eq:phi3split}, by induction on the number of particles. The base case, where the set $A$ is empty ($|A|=0$), is trivial, as the three-point amplitude
${\cal A}^{\phi^3}_{3}(i,j;\kappa)=1$. For the inductive hypothesis, we assume that the $2$-split property, and therefore the associated hidden zero under the constraint of \eqref{eq:2splitcons}, holds for all form factors $F^{{\cal L}_{\phi}}_m$ with fewer than $n$ particles ($m<n$). We prove the property for $F^{{\cal L}_{\phi}}_n$ using the BCFW recursion relation. Under the $2$-split kinematic constraints, the inductive hypothesis implies that most BCFW diagrams vanish. Only the four diagram types shown in Fig. \ref{fig:bcfw} yield non-zero contributions. Each diagram consists of two terms: one where the form factor operator is on the left sub-diagram and the amplitude is on the right, and another with the roles reversed.

\textbf{The first diagram:} In this diagram, the BCFW channel splits the set of particles $A$ into $A_1$ and $\bar{A}_1$.The contribution is:
\begin{align}\label{eq:1stfig}
\sum_{I_1}F^{\cal{L}_\phi}(\hat{i},A_{1},\hat{I}_1;q)\frac{1}{P_{I_{1}}^2}{\cal A}^{\phi^3}(-\hat{I}_1,\bar{A_1},\hat{j},B;k)+{\cal A}^{\phi^3}(\hat{i},A_{1},\hat{I}_1)\frac{1}{P_{I_{1}}^2}F^{\cal{L}_\phi}(-\hat{I}_1,\bar{A_1},\hat{j},B;k,q)\,,
\end{align}
where first term vanishes because the lower-point form factor $F^{\cal{L}_\phi}(\hat{i},A_{1},\hat{I}_1;q)$ is subject to a hidden zero condition from \eqref{eq:ymffzero2}. The second term can be analyzed by applying the $2$-split property to the lower-point form factor inside the sum:
\begin{align}
&\sum_{I_1}{\cal A}^{\phi^3}(\hat{i},A_{1},\hat{I}_1)\frac{1}{P_{I_{1}}^2}\Big({\cal A}^{\phi^3}(-\hat{I}_1,\bar{A_{1}},\hat{j};\kappa_{I_1})F_{|B|+3}^{\cal{L}_\phi}(\hat{j},B,-\hat{I}_1;\kappa_{I_1}',q)\Big)\nn
=&\left(\sum_{I_1}{\cal A}^{\phi^3}(\hat{i},A_{1},\hat{I}_1)\frac{1}{P_{I_{1}}^2}{\cal A}^{\phi^3}(-\hat{I}_1,\bar{A_{1}},\hat{j};\kappa)\right)F_{|B|+3}^{{\cal L}_\phi}(\hat{j},B,\hat{i};\kappa',q)\,.
\end{align}
In the first line,  the momentum conservation determines 
\bea 
p_{\kappa_{I_1}'} =  p_{k}+p_{\bar{A_1}},~~~~ p_{\kappa_{I_1}} =  p_k+p_B-q=p_{\kappa}\,.
\eea
In the second line, we have used a key relation:
\begin{align}
\label{key-observation}
F_{|B|+3}^{{\cal L}_\phi}(\hat{j},B,-\hat{I}_1;\kappa_{I_1}',q)=F_{|B|+3}^{{\cal L}_\phi}(\hat{j},B,\hat{i};\kappa',q)\,.
\end{align}
This equality holds because the propagator along the line connecting $-\hat{I}_1,\hat{j}$ on the left-hand side, dot products with the shifted momentum $\hat{P}_{I_1}$ reduce to dot products with $\hat{p}_{i}$:
\bea 
 \label{observe1}
(-\hat{P}_{I_1}+p_{B'})^2 &= & p_{B'}^2-2\hat{P}_{I_1}\cdot p_{B'}=p_{B'}^2+2(\hat{p}_i+p_{A_{1}})\cdot p_{B'}\nn
& = & p_{B'}^2+2\hat{p}_i\cdot p_{B'}
=(\hat{p}_i+p_{B'})^2\,.
\eea
Here, we have used the condition \eqref{eq:2splitcons} and $\hat{p}_i^2=\hat{P}_{I_1}^2=0$. For the block containing the leg $\kappa_{I_1}'$ on the left-hand side, the inverse of massive propagators is given by 
 \bea 
 (p_{\kappa_{I}'}+p_{B_i})^2- (p_{\kappa_{I}'})^2 &=& p_{B_i}^2+2p_{B_i}\cdot p_{\kappa_{I}'} = p_{B_i}^2+2p_{B_i}\cdot (p_k+p_{A_{1R}}) \nn
 &=& p_{B_i}^2+2p_{B_i}\cdot (p_k+p_{A}) = (p_{\kappa'}+p_{B_i})^2- (p_{\kappa'})^2
 \eea
 Similarly, the numerator factor from the operator $\frac{1}{2}\rm Tr\left((\partial \phi)^2\right)$ is same on the both-hand sides of \eqref{key-observation}:
 \begin{align}
     q\cdot p_{\kappa_{I_1}'}=q\cdot(p_k+p_{\bar{A_1}})=q\cdot p_k\,,\quad -q\cdot \hat{p}_{I_1}=q\cdot(\hat{p}_i+p_{A_1})=q\cdot \hat{p}_i\,.
 \end{align}
 A similar analysis applies to the other three diagrams.
\begin{itemize}
\item \textbf{The second diagram:}
\begin{align}\label{eq:2ndfig}
&\sum_{I_2}F^{{\cal L}_\phi}(\hat{i},A_{2},\hat{I}_2,B(k);q)\frac{1}{P_{I_2}^2}{\cal A}^{\phi^3}(-\hat{I}_2,\bar{A_{2}},\hat{j})\nn
    =&\sum_{I_2}\left({\cal A}^{\phi^3}(\hat{i},A_{2},\hat{I}_2;\kappa_{I_2}) F^{{\cal L}_\phi}(\hat{I},B(\kappa_{I_2}'),\hat{i};q)\right)\frac{1}{P_{I_2}^2}{\cal A}^{\phi^3}(-\hat{I}_2,\bar{A_{2}},\hat{j})\nn
    =&\left(\sum_{I_2}{\cal A}^{\phi^3}(\hat{i},A_{2},\hat{I}_2;\kappa)\frac{1}{P_{I_2}^2}{\cal A}^{\phi^3}(-\hat{I}_2,\bar{A_{2}},\hat{j})\right)F_{|B|+3}^{{\cal L}_\phi}(\hat{j},B(\kappa'),\hat{i};q)\,.
\end{align}
\item \textbf{The third diagram:}
\begin{align}\label{eq:3rdfig}
    &\sum_{I_3}{\scalebox{0.8}{$\Big(F^{{\cal L}_\phi}(B_{3},\hat{i},\hat{I}_3;q)\,{1\over P_{I_3}^2}\,{\cal A}^{\phi^3}(-\hat{I}_3,A,\hat{j},\bar{B}_{3}(k))+{\cal A}^{\phi^3}(B_{3},\hat{i},\hat{I}_3)\,{1\over P_{I_3}^2}\,F^{{\cal L}_\phi}(-\hat{I}_3,A,\hat{j},\bar{B}_{3}(k);q)\Big)$}}\nn
    =&{\cal A}^{\phi^3}(\hat{i},A,\hat{j},B;\kappa)\Big(\sum_{I_3}F^{{\cal L}_\phi}(B_{3},\hat{i},\hat{I}_3;q)\frac{1}{P_{I_3}^2}{\cal A}^{\phi^3}(-\hat{I}_3,\hat{j},\bar{B}_{3};\kappa')\nn
    &~~~~~~~+{\cal A}^{\phi^3}(B_{3},\hat{i},\hat{I}_3)\frac{1}{P_{I_3}^2}F^{{\cal L}_\phi}(-\hat{I}_3,\hat{j},\bar{B}_{3};\kappa',q)\Big)\,.
\end{align}
\item \textbf{The forth diagram:}
\begin{align}\label{eq:4thfig}
    &\sum_{I_4}{\scalebox{0.8}{$\left(F^{{\cal L}_\phi}(B_{4}(k),\hat{i},A,\hat{I}_4;q)\,{1\over P_{I_4}^2}\,{\cal A}^{\phi^3}(\hat{j},\bar{B}_{4},-\hat{I}_4)+{\cal A}^{\phi^3}(B_{4}(k),\hat{i},A,\hat{I}_4)\,{1\over P_{I_4}^2}\,F^{{\cal L}_\phi}(\hat{j},\bar{B}_{4},-\hat{I}_4;q)\right)$}}\nn
    &={\cal A}^{\phi^3}(\hat{i},A,\hat{j},B;\kappa)\Big(\sum_{I_4}F^{{\cal L}_\phi}(B_{4},\hat{i},\hat{I}_4;\kappa',q)\frac{1}{P_{I_4}^2}{\cal A}^{\phi^3}(-\hat{I}_4,A,\hat{j},\bar{B}_{4})\nn
    &~~~~~~~+{\cal A}^{\phi^3}(B_{4},\hat{i},\hat{I}_4;\kappa')\frac{1}{P_{I_4}^2}F^{{\cal L}_\phi}(-\hat{I}_4,\hat{j},\bar{B}_{4};q)\Big)\,.
\end{align}
\end{itemize}
Summing the contributions from these four non-vanishing diagrams in  \eqref{eq:1stfig}, \eqref{eq:2ndfig}, \eqref{eq:3rdfig}, and \eqref{eq:4thfig} precisely reconstructs the BCFW expansion for the right-hand side of the $2$-split formula, \eqref{eq:phi3split}. The first 2 diagrams build the right-hand current, while the last 2 diagrams build the right-hand form factor. This completes the inductive proof for the $2$-split of the form factor $F^{{\cal L}_{\phi}}_n$ \eqref{eq:phi3split}.

\subsection{${\cal O}=\rm Tr(F^2)$}
\label{subsec:ymsplit}

The method to prove the $2$-split for YM form factor is analogous to the scalar case, relying on induction and the BCFW recursion relation. For the set $A$ is empty ($|A|=0$), the case is trivial, as the three-point current
${\cal J}^{\rm mixed}_{3}(i^{\phi},j^{\phi};\kappa^{\phi})=1$. For the inductive hypothesis, we assume that the $2$-split, and therefore the associated hidden zeros \eqref{eq:ymhiddenzero}, holds for all form factors with fewer than $n$ external particles. We prove the property for $F^{\Tr(F^2)}_n$ using the BCFW recursion relation. We apply a BCFW shift to legs $i$ and $j$, choosing the shifted polarizations $\hat{\epsilon}_i$ and $\hat{\epsilon}_j$ to satisfy the constraints of \eqref{eq:polarcon}. Due to the inductive hypothesis, only the four BCFW diagram types previously shown in Fig. \ref{fig:bcfw} give non-zero contributions.

Consider the first diagram, the contribution is:
\begin{align}
    \sum_{I_1}\sum_{h}{\cal A}^{\rm YM}(\hat{i},A_{1},\hat{I}_1){F^{\Tr(F^2)}}(-\hat{I}_1,\bar{A}_{1},\hat{j},B(k);q)\,.
\end{align}
Here, the term where the form factor is on the left sub-diagram, $F^{\Tr(F^2)}(\hat{i},A_{1},\hat{I}_1)$, vanishes due to the polarization constraints in \eqref{eq:polarcon}, so it has been omitted. 

A key step is to resolve the polarization sum. The polarization constraints \eqref{eq:polarcon} require that the shifted polarization $\hat{\epsilon}_i$ can only contract with the polarization of the internal gluon, $\epsilon_I^h$. otherwise the amplitude
${\cal A}^{\rm YM}(\hat{i},A_{1},\hat{I}_1)$ is annihilated due to the condition \eqref{eq:polarcon}. Using $\sum_h\,\big(\epsilon_{I_1}^h\,\epsilon_{-I_1}^{-h}\big)^{\mu\nu}\,\sim\,g^{\mu\nu}$
we can effectively absorb $\hat{\epsilon}_i$ into the form factor by replacing $\epsilon_{-I_1}^{-h}$ with $\hat{\epsilon}_i$. This eliminates the helicity sum and converts the YM amplitude into a YM-Scalar (YMS) amplitude, where legs $\hat{i}$ and $\hat{I}$ are treated as scalars. 
 
Now, applying the inductive hypothesis to the lower-point form factor, we get:
\begin{align}
F^{\Tr(F^2)}(-\hat{I}^{\hat{\epsilon}_i},\bar{A}_{1},\hat{j},B(k);q)\xrightarrow[]{}\,{\cal J}^{\rm mixed} (-\hat{I}^\phi,\bar{A}_{1},\hat{j}^\phi;\kappa_{I_1}^{\phi})F^{\Tr(F^2)}(\hat{j},B,\hat{I}^{\hat{\epsilon}_i};\kappa'_{I_1},q)_{\mu}\epsilon_k^{\mu}\,,
\end{align}
where the superscript $\hat{\epsilon}_i$ is to emphasize that the polarization of the particle $\hat{I}$ is $\hat{\epsilon}_i$.
Thus we arrive at
\begin{align}\label{eq:ymfirst}
&\sum_{I_1}\sum_h\,{\cal A}^{\rm YM}(\hat{i},A_{1},\hat{I}_1)\,{1\over P_{I_1}^2}\,F^{\Tr(F^2)}(-I_1,\bar{A}_{1},\hat{j},B(k);q)\nn
=&\sum_{I_1}{\cal A}^{\rm YMS}(\hat{i}^\phi,A_{1},\hat{I}_1^\phi)\,{1\over P_{I_1}^2}\,{\cal J}^{\rm mixed} (-\hat{I}_1^\phi,\bar{A}_{1},\hat{j}^\phi;\kappa_{I_1}^{\phi})F^{\Tr(F^2)}(\hat{j},B,\hat{I}_1^{\hat{\epsilon}_i};\kappa'_{I_1},q)_{\mu}\epsilon_k^{\mu}\nn
=&\sum_{I_1}{\cal A}^{\rm YMS}(\hat{i}^\phi,A_{1},\hat{I}_1^\phi)\,{1\over P_{I_1}^2}\,{\cal J}^{\rm mixed} (-\hat{I}_1^\phi,\bar{A}_{1},\hat{j}^\phi;\kappa^{\phi})F^{\Tr(F^2)}(\hat{j},B,\hat{i};\kappa',q)_{\mu}\epsilon_k^{\mu}\,,
\end{align}
where $p_{\kappa_{I_1}}=p_B+p_k-q=p_{\kappa}$, $p_{\kappa'_{I_1}}=p_k+p_{\bar{A_1}}$. To get the third equation, we need to prove 
\begin{align}\label{eq:ffeqff}
    F^{\Tr(F^2)}(\hat{j},B,\hat{I}_1^{\hat{\epsilon}_i};\kappa'_{I_1},q)_{\mu}\epsilon_k^{\mu}\overset{?}{=}F^{\Tr(F^2)}(\hat{j},B,\hat{i};\kappa',q)_{\mu}\epsilon_k^{\mu}
\end{align}
Their denominators are identical, as argued in the scalar case \eqref{key-observation}. The numerator is a function of Lorentz contraction. The non-kinematic part from the $\Tr(F^2)$ operator is the same, as argued in \cite{Feng:2025ofq}. The vertex inserted by the kinematic term is $V^{\mu\nu}=p\cdot qg^{\mu\nu}-p^{\mu}q^{\nu}$. Since the polarizations are same in both sides of \eqref{eq:ffeqff} and $q\cdot \{\hat{p}_{I_1},p_{\kappa'_{I_1}}\}=q\cdot \{\hat{p}_i,p_{\kappa'}\}$. Thus,  \eqref{eq:ffeqff} is correct. Analogously, the second diagram is
\begin{align}\label{eq:ymsecond}
&\sum_{I_2}\sum_h\,F^{\Tr(F^2)}(\hat{i},A_{2},\hat{I}_2,B(k);q)\,{1\over P_{I_2}^2}\,{\cal A}^{\rm YM}(-I_2,\bar{A}_{2},\hat{j})\nn
=&\sum_{I_2}{\cal J}^{\rm mixed}(\hat{i}^\phi,A_{2},\hat{I}_2^\phi;\kappa^{\phi})\,{1\over P_{I_2}^2}\,{\cal A}^{\rm YM}(-I_2^{\phi},\bar{A}_{2},\hat{j}^{\phi})F^{\Tr(F^2)}(\hat{j},B,\hat{i};\kappa',q)_{\mu}\epsilon_k^{\mu}\,.
\end{align}
Together, these two diagrams build the right-hand current in the 2-split formula.

For the remaining diagrams, a fully rigorous proof is complex. Instead, we demonstrate the factorization by considering a specific kinematic configuration that satisfies all constraints. We decompose the full kinematic space into two orthogonal subspaces, ${\cal S}={\cal S}_1\oplus{\cal S}_2$. Momenta and polarizations of legs in set $A$ lie in ${\cal S}_1$, while those for legs in set $B$, along with $q$ and polarizations $\epsilon_{i,j,k}$ lie in ${\cal S}_2$. 

Consider the third diagram's contribution:
\begin{align}\label{eq:3ymdiag}
    \sum_{I_3}\sum_{h}&\Big(F^{\Tr(F^2)}(\hat{i},B_3,\hat{I}_3;q)\frac{1}{P_{I_3}^2}{\cal A}^{\rm YM}(-\hat{I}_3,A,\hat{j},\bar{B}_3(k))\nn
    &+{\cal A}^{\rm YM}(\hat{i},B_3,\hat{I}_3)\frac{1}{P_{I_3}^2}F^{\Tr(F^2)}(-\hat{I}_3,A,\hat{j},\bar{B}_3(k);q)\Big)\,.
\end{align}
For the amplitude ${\cal A}^{\rm YM}(\hat{i},B_{3},\hat{I}_3)$ to be non-zero, the effective polarization $\epsilon_{I_3}^H$ must lie in ${\cal S}_1$. This is because all other momenta and polarizations in this amplitude lie in ${\cal S}_1$, except $\hat{p}_i$, lie in ${\cal S}_1$. For $\hat{p}_i$, we can use the momentum conservation $\hat{p}_i=-\hat{P}_{I_3}-p_{B_{3}}$ and the on-shell condition $\epsilon_{I_3}^H\cdot \hat{P}_{I_3}=0$, to confirm $\epsilon^H_{I_3}\cdot\hat{p}_i=0$. Thus, any component of $\epsilon_{I_3}^H$
in ${\cal S}_2$ would lead to a vanishing contraction. 
  
This constraint forces the factorization of the sub-diagrams according to the 2-split hypothesis. The amplitude and the form factor on the right-hand side both split, factoring out the same lower-point current:
\begin{align}\label{eq:3ymdiag1}
&F^{\Tr(F^2)}(-\hat{I}_3, A,\hat{j},\bar{B}_3(k);q)\xrightarrow[]{}\,{\cal J}^{\rm mixed}(-\hat{I}_3^{\phi},A,\hat{j}^{\phi};\kappa^{\phi}_{I_3}){F}^{\Tr(F^2)}(\hat{j},\bar{B}_3,-\hat{I}_{3};\kappa'_{I_3},q)_{\mu}\epsilon^{\mu}_k\,,\nn
&{\cal A}^{\rm YM}(-\hat{I}_3, A,\hat{j},\bar{B}_3(k))\xrightarrow[]{}\,{\cal J}^{\rm mixed}(-\hat{I}_3^{\phi},A,\hat{j}^{\phi};\kappa^{\phi}_{I_3}){\cal J}^{\rm YM}(\hat{j},\bar{B}_3,-\hat{I}_{3};\kappa'_{I_3})\,.
\end{align}
Here, $\kappa'_{I_3}=\kappa'$ follow from momentum conservation. From the scalar case discussion and \cite{Feng:2025ofq}, we know 
\begin{align}\label{eq:3ymdiag3}
    {\cal J}^{\rm mixed}(-\hat{I}_3^{\phi},A,\hat{j}^{\phi};\kappa^{\phi}_{I_3})={\cal J}^{\rm mixed}(i^{\phi},A,\hat{j}^{\phi};\kappa^{\phi})
\end{align}
Plugging \eqref{eq:3ymdiag1}, and \eqref{eq:3ymdiag3} into \eqref{eq:3ymdiag}, we can factor out this common current:
\begin{align}\label{eq:ymthird}
    {\cal J}^{\rm mixed}(i^{\phi},A,\hat{j}^{\phi};\kappa^{\phi})\sum_{I_3}\sum_{h}&\Big(F^{\Tr(F^2)}(\hat{i},B_3,\hat{I}_3;q)\frac{1}{P_{I_3}^2}{\cal J}^{\rm YM}(-\hat{I}_3,\hat{j},\bar{B}_3;\kappa')\nn
    &+{\cal A}^{\rm YM}(\hat{i},B_3,\hat{I}_3)\frac{1}{P_{I_3}^2}F^{\Tr(F^2)}(-\hat{I}_3,\hat{j},\bar{B}_3;\kappa',q)\Big)
\end{align}
The same logic applies to the fourth diagram:
\begin{align}\label{eq:ymfourth}
    {\cal J}^{\rm mixed}(i^{\phi},A,\hat{j}^{\phi};\kappa^{\phi})\sum_{I_3}\sum_{h}&\Big(F^{\Tr(F^2)}(\hat{i},B_3,\hat{I}_3;\kappa',q)\frac{1}{P_{I_3}^2}{\cal J}^{\rm YM}(-\hat{I}_3,\hat{j},\bar{B}_3)\nn
    &+{\cal A}^{\rm YM}(\hat{i},B_3,\hat{I}_3;\kappa')\frac{1}{P_{I_3}^2}F^{\Tr(F^2)}(-\hat{I}_3,\hat{j},\bar{B}_3;q)\Big)
\end{align}
The sums in the brackets for the third and forth diagrams then combine to form the BCFW construction of the right-hand form factor. 

Therefore, combining Eqs.~\eqref{eq:ymfirst}, \eqref{eq:ymsecond}, \eqref{eq:ymthird}, and \eqref{eq:ymfourth} yields precisely the BCFW recursion relation for the right-hand side of \eqref{eq:ymsplit}.

\section{\label{sec:con}Conclusion}

In this work, our central result is the proof of the $2$-split factorization for the form factors of the operators ${\cal O} = \tfrac{1}{2}\Tr((\partial\phi)^2) + \Tr(\phi^3)$ and ${\cal O} = \Tr(F^2)$. By imposing specific kinematic constraints that partition the external legs, \cite{Cao:2025hio} showed that an $n$-point form factor factorizes into a product of a lower-point amplitude and a lower-point form factor. We prove this behavior by induction. The base cases were established directly, and the inductive step was carried out by applying the BCFW recursion relation. We showed that under the $2$-split kinematic conditions, only a select few BCFW diagrams contribute, and their sum precisely reconstructs the factorized form. A crucial element of our analysis was the identification of hidden zeros, which are necessary for the consistency of the inductive proof. Unlike the proof for amplitudes in \cite{Feng:2025ofq}, these hidden zeros are insufficient to establish the $2$-split property for form factors. An additional zero is required, ensuring the vanishing of $F(\hat{i},A,\hat{I};q)$ (and $F(\hat{I},A,\hat{j};q)$).

The techniques and results presented here further bridge the gap between the study of scattering amplitudes and form factors, highlighting the universal applicability of on-shell methods. This opens several avenues for future research. An obvious next step would be to investigate whether similar factorization properties hold for form factors of operators in effective field theories or theories with massive particles would also be a valuable direction. Ultimately, it remains an open question whether the form factor can be bootstrapped from its zeros, as in \cite{Rodina:2024yfc}.

\section*{Acknowledgments}

We thank Qu Cao for valuable discussions, and Bo Feng and Kang Zhou for helpful comments on the manuscript. This work was supported by the CSRC Postdoctoral Budget.

\bibliography{reference}
\bibliographystyle{JHEP}

\end{document}